\documentclass[referee]{aa} 
\usepackage{graphicx}
\usepackage{txfonts}
\usepackage[figuresright]{rotating}
\begin{document}
   \title{Spherical Accretion in Nearby Weakly Active Galaxies}

   \author{M. Moscibrodzka
          \inst{1}
          }

   \offprints{mmosc@camk.edu.pl}

   \institute{1. N. Copernicus Astronomical Center, 
              Bartycka 18, 00-716 ,Warsaw ,Poland\\
                          }

   \date{Received 07 September 2005 /Accepted 15 December 2005}

   \abstract{ We consider the sample of weakly
 active galaxies situated in 'Local Universe' collected in the paper of Pellegrini (2005) with inferred accretion efficiencies from $10^{-2}$ to $10^{-7}$.
 We apply a model of spherically symmetrical Bondi accretion for given 
parameters ($M_{BH}$,$T_{\infty}$,$\rho_{\infty}$,) 
 taken from observation. We calculate spectra emitted by the gas accreting onto 
its central objects using Monte Carlo method including synchrotron and 
bremsstrahlung photons as seed photons. 
 We compare our results with observed  nuclear X-ray luminosities $L_{X,nuc}$ (0.3-10 keV) of the sample.
Model is also tested for different external medium parameters ($\rho_{\infty}$ and $T_{\infty}$) and different free parameters of the model. 
Our model is able to explain observed nuclear luminosities $L_X$ under an assumption that half of the compresion energy is transfered directly
to the electrons. 

   }

   \maketitle
%

\section{Introduction}
It is believed that most of the galactic nuclei host supper-massive black holes in their centers.
The masses of these central objects estimated by various methods are of the order of $10^6-10^9 \rm{M_{\odot}}$.
The radiation emerging from the vicinity of these objects is produced by an accretion process. 
We believe that Quasars and Seyfert galaxies are powered by the flows with high
 angular momentum and are characterized by high accretion efficiency. Corresponding models 
consisting of thin accretion disk surrounded by some kind of a  
corona are able to explain
all spectral features like big blue bump, iron line and short time-scale variability.
The accretion luminosity in these sources is of order of a few percent of 
the Eddington limit or more. Nevertheless there exist a number ($40\%$ of nearby galaxies, Ho 2003) of 
galactic nuclei which are very faint objects 
although they contain black holes of masses of the same order as those in AGN. 
For example Galactic Center hosts a super-massive object of mass $\sim 10^6 \rm{M_{\odot}}$ (Genzel et al. 2003).
The X-ray luminosity of this object is about $10^{-9}$ times smaller than its Eddington limit.
The density and the temperature near the capture radius were estimated from Chandra data (Baganoff et al. 2003), 
which allowed to compute the expected mass accretion rate, and calculate the dynamics 
of the accreting gas using for example the simplest Bondi model (1952) 
of the steady spherically symmetric flow. The efficiency of this accretion flow, as estimated from
comparison of expected accretion rates and observed luminosities, is about $\eta= 10^{-5}$,
 which is very small in comparison to the typical active quasar value where efficiency is $\eta=0.1$.
Galactic Center is an example of the extremely faint source, sometimes it is even called inactive (Nayakshin 2004). 
In paper of Pellegrini (2005, hereafter P2005) there are collected more examples of faint galactic 
nuclei sources (LLAGN, Low Luminosity Active Galactic Nuclei) located in the 'Local Universe'.
The masses of black holes in centers of these objects obtained by various methods (for details see P2005)
lay in the same range as typical $M_{BH}$ AGN. Their luminosities are of the order of $10^{-2}-10^{-7}$ of  
the Eddington luminosity. The efficiency of the accretion (assuming Bondi mass accretion rate) is thus also low. 
LLAGN have no big blue bump (Quataert et al. 1999), weak iron 
lines and no short time-scale variability (Ptak et al. 1998) and this is an evidence of nonexistence  
of the accretion disk inside the flow (Nayakshin 2004 and references therein).
To build the consistent picture we would like those objects to be quiescent phases 
of long-term evolution of quasars (Nayakshin 2004).
To explain such a low activity usually we assume one of the RIAF (radiatively inefficient accretion flow) solutions. 
Examples of RIAFs include Bondi
 flow with no angular momentum, low angular momentum flows (Proga 2003) and ADAF with high angular momentum (Narayan 2002). 
Other possibilities are spherical flows with magnetic fields,
convection dominated flows (Narayan 2002 and references therein) or jet-wind accretion flows (Yuan 2003).\\
In this paper, to explain low luminosities of the sample of LLAGN,
 we apply the spherically symmetrical accretion in the Newtonian regime (Bondi 1952).
In this model accretion rate is determined by external medium conditions and remain constant
in the whole radius range, thus in this model we do not include an outflow in any form.
 Bondi flow can be also treated as rough approximation of very low angular momentum flows in a steady phase. 
Plasma accreting onto a central object in our model is a source of synchrotron and bremsstrahlung photons.
We assume that plasma has a two-temperature structure. Because in our model cooling of the flow occur by electrons,
ions are much hotter than electrons near the horizon.    
In calculations of emerging radiation spectrum we include
the density and the temperatures of ions and electrons as a function of radius.
Radiative transfer of synchrotron and bremsstrahlung radiation through the plasma is calculated using 
Monte Carlo technique (Gorecki $\&$ Wilczewski 1984, Pozdnyakov et al.1983). Because of 
the method of calculating the spectra, we neglect the radiation pressure 
influence on the flow dynamics.

In Sect. 2 we describe the model and the technique of 
calculating the comptonization in details. We present the result in Sect. 3.
Discussion is given in Sect. 4. 


\section{Model}

\subsection{Dynamics of the flow}
The dynamics of the flow is given by Bondi solution (Bondi 1952). We assume the black hole mass $M_{BH}$, 
density of the matter at the infinity $\rho_{\infty}$ and temperature at infinity $T_{\infty}$ to calculate 
the density of matter and velocity of infall profile. No additional sources of the gas or outflow are included. 
The polytropic index $\gamma$ is assumed to be equal close to $\frac{5}{3}$. 
This is a reasonable assumption since
 matter in the close vicinity of a black hole is fully ionized. 
The Bondi radius is then very close to a black hole, of the order of a few Schwarzschild radii.
The flow is then subsonic in almost whole range. 

\subsection{Temperature of the electrons and ions}
To calculate the ion and electron temperature we use the iterative procedure given by Esin (et al.1997).
At first we assume a reasonable electron temperature $T_e$ profile 
and calculate the energetic balance at each point of the flow. The ion temperature is 
determined simultaneously with the corrections to the electron temperature $T_e$ from relation:
\begin{equation}
P_{gas}=\beta \rho c_s^2=n_i k_b T_i + n_e k_b T_e
\end{equation}
where $\beta$ is gas to total pressure ratio, $n_e$ and $n_i$ are respectively electrons and ions 
total numbers per $cm^3$.
The condition of the thermal equilibrium for electrons is given by:
\begin{equation}
Q_{adv}=Q^{ie}+\delta Q_{grav}-Q^{rad}
\label{eq:bilans}
\end{equation}
On the left hand side we have the advection term.
 On the right there is respectively 
heating rate of the electrons by a Coulomb coupling, accretion energy (compression heating) multiplied
 by a small $\delta$ parameter, and a cooling rate.
The heating rate due to the Coulomb 
coupling between ions and electrons is given by (e.g. Rozanska et al. 2000):
\begin{equation}
q^{ie}= 1.5 k_b m_H ln \Lambda \rho^2 2.44 \times 10^{21} T_e^{-1.5} (T_p-T_e).
\label{eq:bilans}
\end{equation}
Cooling rate $Q^{rad}$ consists of four terms: the synchrotron cooling and 
comptonization of synchrotron photons cooling, 
bremsstrahlung cooling and comptonization of bremsstrahlung. 
\begin{equation}
Q^{rad} = q^-_{S}+q^-_{C,S}+q^-_{brem}+q^-_{C,brem}
\end{equation}
 Derivation of $q^-_{C,S}$ and $q^-_{C,brem}$ is described in Esin et al.(1996).
The cooling by the synchrotron photons is given by:
\begin{equation}
q_S^- = \int_0^{\infty} \epsilon_S (\nu) e^{-\tau_{\nu}}  d\nu
\end{equation}    
The exponential term is due to a strong selfabsorption of the 
synchrotron photons below certain frequency. The optical depth 
is a function of frequency and we can calculate it by estimating the escape probability 
of the photon of given energy from a given location. 
To find this probability we follow N trajectories of
 photons (rays) with randomly chosen directions. We follow the rays 
along and calculate the optical depth. The optical depth
 is given by:
\begin{equation}
\tau_{\nu} = \int_0^{r_{boundary}} \alpha_{\nu}^S dl = \int_0^{r_{boundary}} \frac{\epsilon_S(\nu)}{4 \pi B_{T_e}(\nu)} dl
\label{eq:tau}
\end{equation}
Here $\alpha_{\nu}^S$ is the absorption coefficient. The second part of the Eq.~\ref{eq:tau} is a Kirchoff's law.
The mean optical depth is :
\begin{equation}
\hat{\tau}_{\nu}= \frac{1}{N} \sum_{i=1}^{N}e^{-\tau^i_{\nu}}
\end{equation}
This procedure of deriving $q^-$ is described in more details 
in Kurpiewski et al. (2001).

In thermal balance equation we assume the part of the energy must be advected 
without emission. Term $Q_{adv}$ is calculated as described e.g. in Narayan et al. (1998). 
We also assume that part of the compression energy taken by ions is
 taken by the electrons. This is given by $\delta Q_{grav}$ where
 $\delta$ in a parameter of this model. This energy is proportional to the temperature of 
ions, the density, and the velocity of the falling matter ($\rm{v=v_r}$). It is given by:
\begin{equation}
Q_{grav}=\frac{3}{2} \frac{k_b T_i \rho}{\mu m_H r} \rm{v}
\end{equation}

Corrections to the electron temperature are calculated from energy balance equation iteratively,  
until they are less that $1\%$ of the $T_e$.   

\subsection{Emissivity of plasma}
We take into account bremsstrahlung 
photons and synchrotron photons. The emissivity of synchrotron is given
 by Pacholczyk (1970) formula:
\begin{equation}
\epsilon_S(\nu)=4.43 \times 10^{-30} \frac{4 \pi \nu n_e}{K_2(1/\theta)} I(\frac{x_M}{\sin \Phi})  \rm{[ergs/s/cm^3/Hz]}
\end{equation}
where: 
\begin{equation}
x_M=\frac{2 \nu}{3 \nu_0 \theta_2}   \;\;\;\;   \nu_0=\frac{e B}{2 \pi m_e c}
\end{equation}
$\theta$ is dimensionless electron temperature, B is the magnetic field, 
$\nu_0$ is the cyclotron frequency. $\Phi$ is the
 angle between the velocity vector and the direction of local magnetic
 field. The B field is calculated assuming beta parameter defined as $\beta=P_{gas}/P_{tot}$. 
For isotropic velocity function we can average $I(\frac{x_M}{\Phi})$
 over $\Phi$ angle. The averaged function was given by Mahadevan et al. (1996):
\begin{equation}
I'(x_M)=\frac{4.0505}{x_M^{1/6}} (1+\frac{0.40}{x_M^{1/4}}+\frac{0.5316}{x_M^{1/2}}) exp(-1.8899 x_M^{1/3})
\label{eq:I}
\end{equation}
Because the approximation above is good for ultra-relativistic electrons, 
additionally for lower temperatures ($ T_e <10^{10} K$)
we use the modified Eq:~\ref{eq:I}. It takes the form:
\begin{equation}
I'(x_M)=\frac{4.0505 \alpha'}{x_M^{1/6}} (1+\frac{0.40 \beta'}{x_M^{1/4}}+\frac{0.5316 \gamma'}{x_M^{1/2}}) exp(-1.8899 x_M^{1/3})
\label{eq:I2}
\end{equation}  
The coefficients are derived from comparing the synchrotron emissivity 
with the cyclo-synchrotron
emissivity. We take this coefficients from Mahadevan (et al. 1996), 
and interpolate them. For ultra-relativistic temperatures the $\alpha'$,$\beta'$ and $\gamma'$
coefficients approach to 1.
The self-absorption of the synchrotron photons 
is described in previous subsection. 
The bremsstrahlung emissivity is given by:
\begin{equation}
\epsilon^{ff}_{\nu}= q^-_{br} e^{-h\nu/k_bT} \bar{g}_{ff} \rm{[ergs/s/cm^3/Hz]}
\end{equation}
where $q^-_{br}=q_{ie}+q^-_{ee}$.$q_{ie}$ is for ion-electron interaction and $q_{ee}$ is for electron-electron interaction.
Electrons are radiating in these two processes. The formulas for these emissivities are standard taken from e.g. Narayan (et al.1995).
We also include the absorption of bremsstrahlung emission which is calculated in the same manner
 as for synchrotron radiation.

\subsection{Photon generation}
The Monte Carlo method given by Gorecki $\&$ Wilczewski (1984) is based on following 
the photons (or rays) trajectories. At first we generate some initial 
values of the photon energy, place of birth and direction in which the photon
 is emitted.
\begin {itemize}
\item {\bf photon energy.} The photon energy is randomly generated from the
 photon distribution in given location. The photon distribution is given by
 the energy distribution divided by $h \nu$ and additionally multiplied by 
the $exp(-\tau_{\nu})$ which indicates the photon self-absorption. The photon
 distribution is given by:
\begin{equation}
f_{i,j,l}=\frac{n_{\nu}(r_i,\theta_j,\phi_l)}{\int_0^{\infty} n_{\nu}(r_i,\theta_j,\phi_l) d\nu}
\end{equation}
i,j,l numbers indicate the cell numbers (see Sect. 2.7), they correspond respectively to
 radius r,$\theta$ and $\phi$ angle.
\begin{equation}
n_{\nu}(r_i,\theta_j,\phi_l)=\frac{\epsilon_{\nu}(r_i,\theta_j,\phi_l) e^{-\tau_{\nu}}}{h \nu} 
\end{equation}
The initial photon energy $h \nu_0$ is determined by the inversion of 
the cumulative distribution function.
\item {\bf initial position vector.} The initial position of the emission 
of the photon is given by the distribution that is integrated photon 
distribution from zero to infinity in a frequency.
\begin{equation}
f_{i,j,k}=\frac{\dot{N}_{i,j,l}}{\sum_{i,j,l} \dot{N}_{i,j,l}}
\end{equation}
\begin{equation}
\dot{N}_{i,j,l}=\Delta V_{i,j,l} \int_0^{\infty} \frac{\epsilon(\nu) e^{-\hat{\tau}_{\nu}}}{h \nu} d\nu 
\end{equation}
where $\Delta V_{i,j,l}$ is the cell volume. The angles $\cos \theta$ 
and $\phi$ are generated from uniform distributions in ranges 
$\cos \theta \in (-1,1)$ and $\phi \in (0,2 \pi)$.  
\item {\bf direction of emission.} The direction of emission is 
randomly chosen from uniform distributions. The direction is given
 by a vector  ${\bf \Omega } = (\sin \Theta \cos \phi,\sin \Theta \sin \phi,\cos \Theta)$,
 where: $\cos \Theta \in [-1,1]$ and $\phi \in [0;2 \pi]$. 

\item {\bf weight of the ray.} The initial weight is $w_0=1$ , 
and we follow the trajectory of a ray until $w=w_{min}$. In our calculations we adopt 
$w_{min} =10^{-7}$ (Kurpiewski et al 2001).
\end{itemize}

\subsection{Photon propagation}
To count the spectrum we follow the method of Gorecki $\&$ Wilczewski.
 Each ray initially has a weight of $w_0=1$, energy $h\nu_0$, direction 
{\bf$\Omega_0$} and it is emitted from ${\bf r}_0$ location. We construct 
vector $T_i=({\bf r}_i,{\bf \Omega}_i,h\nu_i,w_i)$, where $T_0=({\bf r}_0,{\bf \Omega}_0,h\nu_0,w_0)$. 
Knowing the zero order values we can perform the next step. To determine
  the emerging spectrum we calculate the probability that the photon
 leaves the cloud, which it is given by:
\begin{equation}
P(l_i)=exp(-\int_0^{l_i} N_e <\sigma> dl )
\end{equation}
where $l_i$ is the distance to the boundary of the cloud from a given 
location (a place of last scattering) in a chosen direction. $N_e$ is the
 electron density and $<\sigma>$ is the cross section for scattering, and
 it is given by:
\begin{equation}
<\sigma>=\frac{1}{N_e} \int N(\vec{v}) (1-\vec{v} \cdot \vec{\Omega}/c) \sigma (x) d^3 v
\label{eq:sigma}
\end{equation}
 The cross section is averaged additionally by the term which 
takes into account the probability of the photon - electron 
interaction. To obtain emerging spectrum we are counting $P_iw_i$ in energy
 bands. The next value of $w_{i+1}=w_i(P_i-1)$. To obtain the position of 
the next scattering we model the mean free path distribution. It is given by:
\begin{equation}
\rho(\lambda_i)=\frac{\frac{d\tau(\lambda_i)}{d\lambda_i} e^{-\tau(\lambda_i)}} {\int_0^{\lambda_i} \frac{d\tau(\lambda_i)}{d\lambda_i} e^{-\tau(\lambda_i)}  d\lambda }
\end{equation}
where:
\begin{equation}
\tau(\lambda)=\int_0^{\lambda} Ne <\sigma> dx
\end{equation}
The next collision vector is given by:
\begin{equation}
{\bf \vec{r}}_{i+1} =  {\bf \vec{r}}_{i} + \lambda_i {\bf \vec{\Omega}}_i
\end{equation}
The ${\bf \Omega}_{i+1} $  vector is generated from uniform distributions 
of $\cos \Theta$ and $\phi$. The energy of a new photon after collision is given
 by standard formulas. (For details, how to model the distributions, see
 Gorecki $\&$ Wilczewski 1984).   

\subsection{Seed photons}
To include both synchrotron and bremsstrahlung emission we count
  both contributions to spectra separately. After that we add spectra to each other 
with proper weights. The total number of emitted synchrotron photons 
is given by:
\begin{equation}
N_S=\int \frac{\epsilon_S(\nu)}{h \nu} e^{-\tau_{synch}}d\nu
\end{equation}
and total number of bremsstrahlung photons produced in the flow is:
\begin{equation}
N_B=\int \frac{\epsilon_B(\nu)}{h \nu} e^{-\tau_{brem}}d\nu
\end{equation}
We calculate this integrals and we obtain the weight of the bremsstrahlung
 photons which is given by:
\begin{equation}
\eta_{B-S}=\frac{N_B}{N_S}
\end{equation}
To obtain the whole spectrum we 
add the synchrotron spectrum to the bremsstrahlung spectrum multiplied by factor $\eta_{B-S}$.

\subsection{Numerical calculations and parameters of the model}
In our calculations we divide the flow into cells. 
Geometry of the flow is represented in spherical coordinates.
 We divide the flow in radius in logarithmic scale and angles $\cos \Theta$ and $\phi$. 
The flow is divided into a $10^6$ 
($100 \times 100 \times 100$) cells.  We assume that the value in the middle 
of the cell of the quantities is the mean value in the whole cell. 
The code was tested with homogeneous 
plasma, with central source of black body photons, and also for ADAF
solutions. It gave similar results as in other papers of Gorecki $\&$ Wilczewski (1984) and Narayan (et al. 1995).
The difference between Narayan (et al. 1995) and our approach is that we include the self-absorption of bremsstrahlung photons.
Also the results are slightly different because in ADAF the authors use different geometry.
The code also can be used for any dynamical solution.
Parameters of the model are: the black hole mass $M_{BH}$, initial values at infinity $\rho_{\infty}$, $T_{\infty}$, and $\gamma$. 
First three parameters
are taken from observations of a given source. $\gamma$ is been assumed to be close to $5/3$ which is good for the fully ionized 
plasma (in calculations $\gamma=1.666$, because of numerical difficulties). $\beta$ and $\delta$ are free parameters that influence the electron temperature.
Outer boundary of the spherical cloud is chosen to be Bondi radius. 
The boundary of the flow is also a parameter of the model. 
The inner radius of comptonizing cloud is always fixed at the 3 Schwarzschild radii. 
 
\section{Results}
\subsection{Model parameter tests}
In general the radiation spectra 
emitted from the vicinity of the central objects in our model consist of two components. First one is created by synchrotron photons, 
the second one by bremsstrahlung photons. Most of the 
synchrotron photons come out from the spherical cloud of plasma without scattering and create a bump in radio band.
Synchrotron bump in some cases can be shifted to IR band depending on the parameters of the model ($M_{BH}$,$\rho_{\infty}$,$T_{\infty}$,$\beta$,$\delta$).
The second weaker bump in radio-IR-opt-UV range is created by synchrotron photons that run out from the spherical cloud of plasma 
after first scattering. The frequency range in which the bump, created by the single scattered photons, is exactly located 
depends also on parameters of the model. Because of the low optical depth (in all cases $\tau < 1$) of the accreting gas, 
the bump created by photons after first scattering is lower than main synchrotron peak. Probability that photons undergoes 
k-th scattering before escaping the cloud is proportional to $\tau^k$. 
The X-$\gamma$ ray bump is mostly created by bremsstrahlung photons. It can be affected by the synchrotron photons 
scattered more than once, if the plasma optical thickness is high enough.

\subsubsection{Results for different $\beta$ parameters}
\begin{figure}
  \centering
  \includegraphics[width=0.5\textwidth]{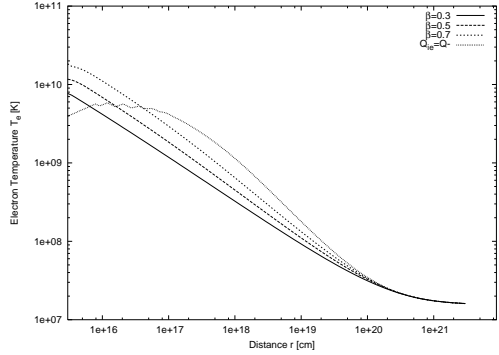}
  \caption{ Electron temperature profiles for different model parameters $\beta$, for fixed $\delta$=0.001.}
  \label{Fig:1}
\end{figure}

\begin{figure}
  \centering
  \includegraphics[width=0.5\textwidth]{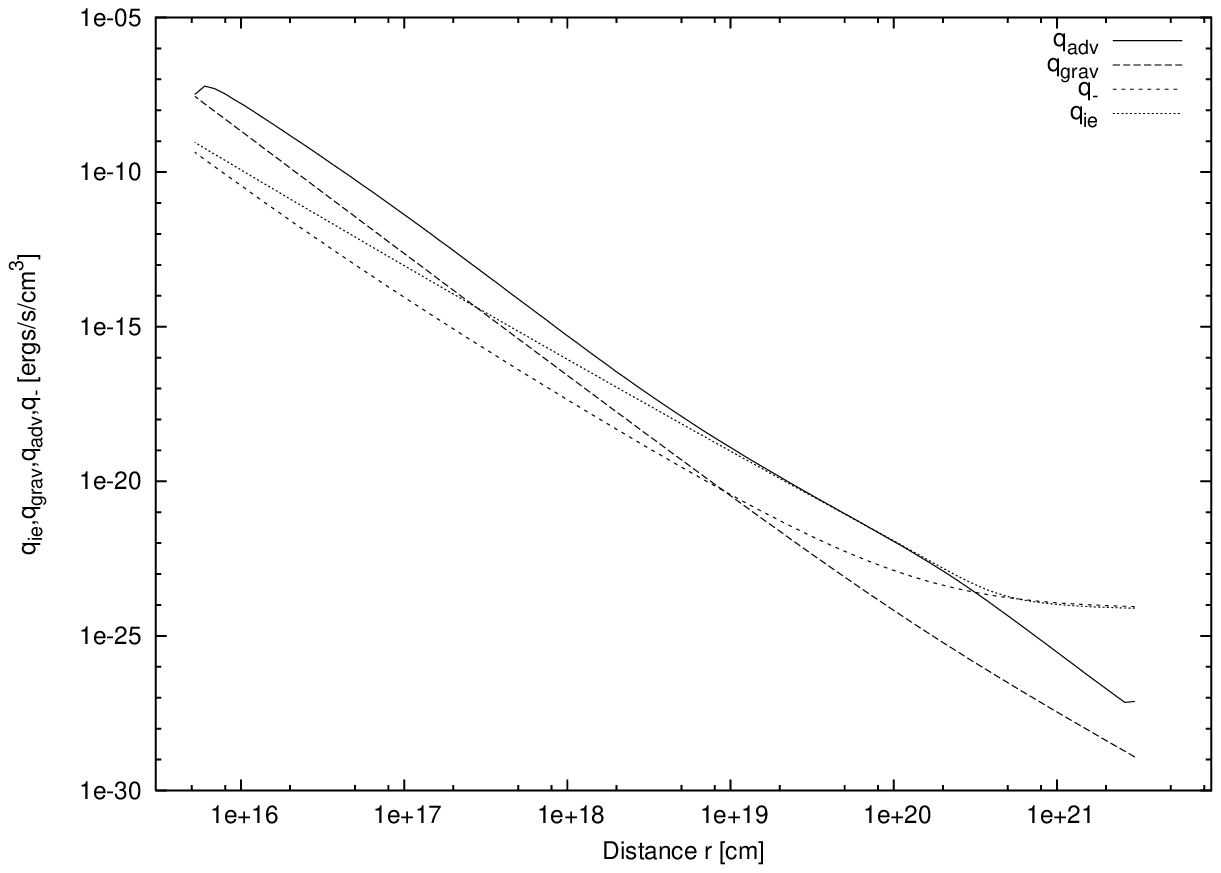}
  \caption{ Heating and cooling terms as a function of radius for parameter $\delta$=0.001 and $\beta$=0.5.}
  \label{Fig:2}
\end{figure}

\begin{figure}
  \centering
  \includegraphics[width=0.5\textwidth]{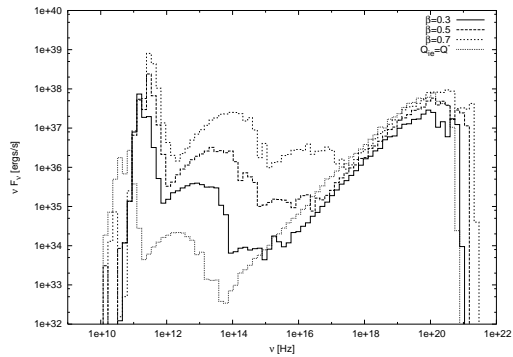}
  \caption{ Calculated spectra of M87 for model parameters $\beta= 0.3,0.5,0.7$ }
  \label{Fig:5}
\end{figure}

To test model for different parameters in details, we performed calculations in case of M87. The electron temperature profiles for different $\beta$
values ($\beta$=0.3,0.5,0.7) for this source are presented in Fig:~\ref{Fig:1}. We assume that electrons receive only a small fraction of compression energy
and we set $\delta$=0.001 (because of the ratio $m_e/m_p$, in our basic model we assume that compression energy is mostly taken by ions because 
they are more massive than electrons, we also perform calculations for other values of $\delta$). 
For comparison in Fig~\ref{Fig:1} we also plot electron temperature profile obtained by iterating 
Eq:~\ref{eq:bilans} without advection $Q_{adv}$ and compression heating $\delta Q_{grav}$ terms, marked as $Q_{ie}=Q^{-}$.  In Fig:~\ref{Fig:2} we present
how four terms of heating ($Q_{ie},\delta Q_{grav}$) and cooling ($Q^{-},Q_{adv}$) are balancing each other for $\beta$=0.5 and $\delta$=0.001.
Far away from a horizon the dominating term in Eq:~\ref{eq:bilans}  are electron-ion coupling and radiation cooling term, the electron temperature
is there the same as in case when we neglect advection and compression heating terms. Close to the black hole the electron temperature is
determined by advection term, which is dominating in this region. Advection term itself consist of two parts which are balancing each other .
The other terms are negligible near the black hole. Electrons reach higher temperatures close to the horizon 
than in case when we balance only $Q_{ie}$ and $Q^{-}$. On the other hand in case when energy balance equation is: $Q_{ie}=Q^{-}$ 
electrons are hotter further from the black hole.
The problem of iterating the electron temperature in case when the advection term is dominating, becomes local. The non-locality caused 
by the radiative cooling term $Q^{-}$ is now negligible. An increase of $\beta$ parameter causes the decrease of magnetic field and 
to increase of the electron temperature.  
Spectra calculated for temperature profiles presented in Fig:~\ref{Fig:1} are plotted
in Fig:~\ref{Fig:5}. The calculations were made for boundary radius of the cloud equal to Bondi radius.
The difference among these cases concerns mainly the synchrotron part of the spectrum. The synchrotron emissivity increases
for higher $\beta$ values.
This is because most of the synchrotron photons are
created very close to the black hole where the temperatures differ the most. The $L_X$ luminosities resulting from  integrating the
outgoing radiation spectrum in range 0.3-10 keV, do not depend strongly from the $\beta$ parameter. X-ray band of the spectrum 
is dominated by the bremsstrahlung photons created respectively further form a black hole. Calculated $log_10(L_X)$ for $\beta$=0.3,0.5,0.7 are
respectively 37.0,37.15,37.42, where $L_X$ is meassured in ergs/s. In case of $\beta$=0.7 the X-ray band is affected a little by synchrotron photons scattered in the medium
more than once. For $Q_{ie}=Q^{-}$ case, synchrotron emissivity is much lower. This is because the temperature near the horizon is smaller. 
$log_{10}(L_X)$=37.44 in this case is larger because at larger distanced electron temperature is higher.

\subsubsection{Results for different $\delta$ parameters}
\begin{figure}
  \centering
  \includegraphics[width=0.5\textwidth]{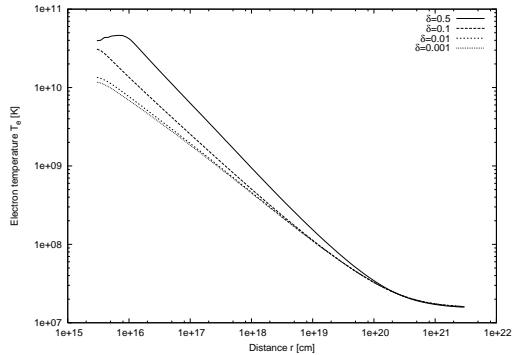}
  \caption{ Electron temperature profiles for different model parameters $\delta$ with fixed $\beta=$0.5.}
  \label{Fig:3}
\end{figure}

\begin{figure}
  \centering
  \includegraphics[width=0.5\textwidth]{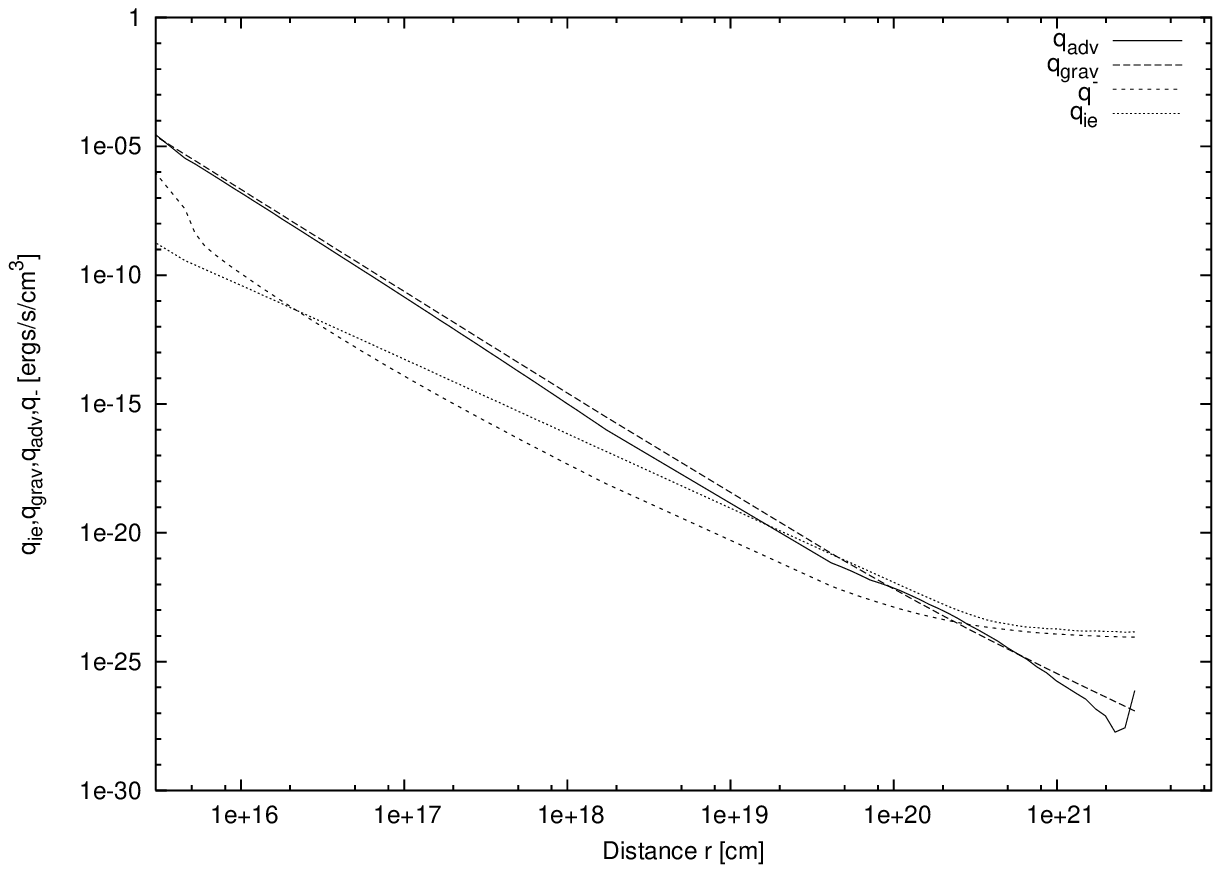}
  \caption{ Heating and cooling terms as a function of radius for parameter $\delta$=0.1 and $\beta$=0.5}
  \label{Fig:4}
\end{figure}

\begin{figure}
  \centering
  \includegraphics[width=0.5\textwidth]{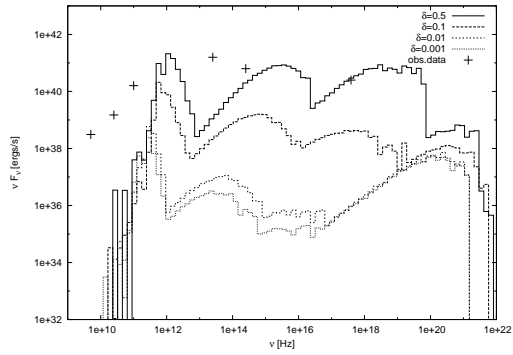}
  \caption{ Calculated spectra of M87 for model parameters  $\delta= 0.001, 0.01, 0.1, 0.5$.}
  \label{Fig:6}
\end{figure}

Since the representative value of $\delta$ is disputable (e.g. Yuan, Narayan $\&$ Yi 1995, Bisnovatui-Kogan $\&$ Lovelace 1997, Narayan $\& $Yi 2003) 
in Fig~\ref{Fig:3} we show the effect of changing $\delta$ parameter in Eq:~\ref{eq:bilans}. We made calculations for four values of $\delta$,
namely 0.001, 0.01, 0.1 and 0.5. $\beta$ is set to be 0.5. For larger $\delta$ more compression energy is put into electrons. The largest difference
between electron temperature profiles are for $\delta=0.1$ and for $\delta=0.5$. Heating and
cooling terms as a function of radius for $\delta=0.1$ are shown in Fig:~\ref{Fig:4}. In this case advection terms is not dominating in the
central parts of the flow. This time $Q_{adv}$ is balanced by compression heating term $\delta Q_{grav}$. In outer part of the flow $Q_{ie}$ and $Q^{-}$
are balancing each other, the same as for $\delta=0.001$. The temperature converge to the same values at infinity.
 Radiation spectra for $\delta$=0.001, 0.01, 0.1 and 0.5 are presented in Fig:~\ref{Fig:6}.
The trend is similar as in Fig:~\ref{Fig:5}. For larger $\delta$ hotter electrons emits more synchrotron photons from the central parts of the flow.
Bremsstrahlung emissivity in the spectrum is slightly affected by changing $\delta$ parameter. The differences in $log_{10}(L_X)$ which is for 
$\delta$=0.001, 0.01, 0.1, 0.5 respectively 
37.15, 37.2, 39.55, 41.15 is caused mainly by the fact that X-ray range 
is dominated by synchrotron photons scattered more than once. 
\subsubsection{Results for different external medium conditions}
\begin{figure}
  \centering
  \includegraphics[angle=0,width=0.5\textwidth]{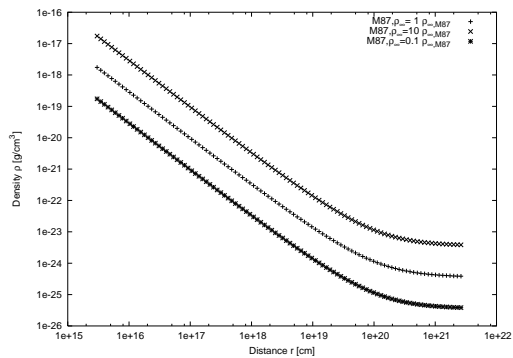}
  \caption{ Density profiles for different initial $\rho_{\infty}$. Calculations made for M87.}
  \label{Fig:7}
\end{figure}

\begin{figure}
  \centering
  \includegraphics[angle=0,width=0.5\textwidth]{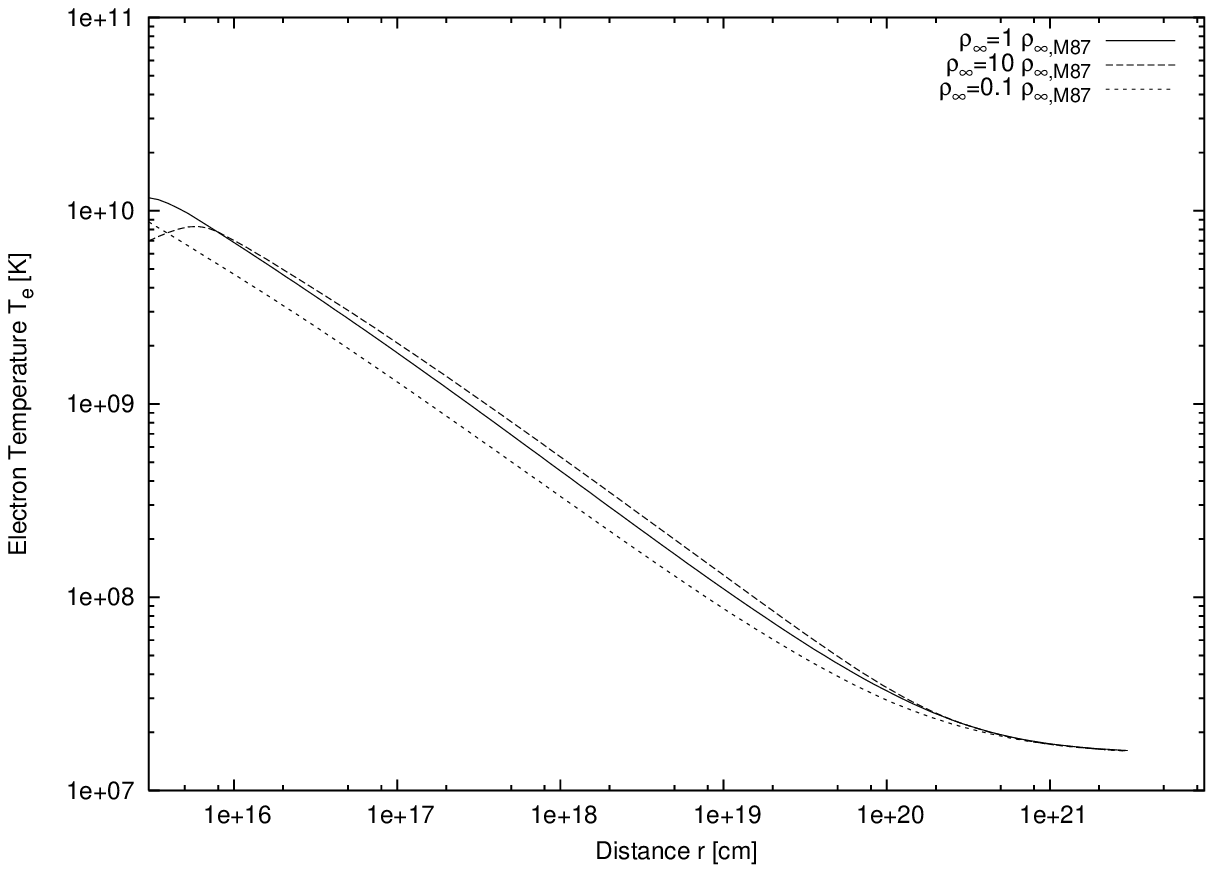}
  \caption{ Electron temperature profiles for different initial $\rho_{\infty}$. Calculations made for M87.}
  \label{Fig:8}
\end{figure}

\begin{figure}
  \centering
  \includegraphics[angle=0,width=0.5\textwidth]{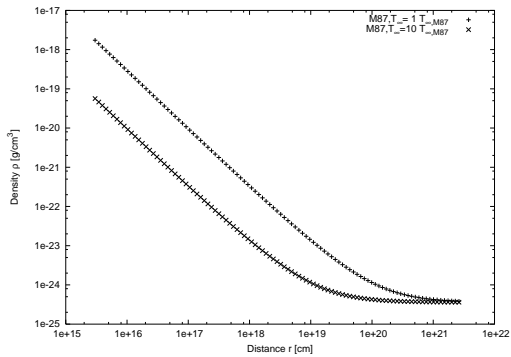}
  \caption{Density profiles for different initial $k_B T_{\infty}$. Calculations made for M87. }
  \label{Fig:9}
\end{figure}

\begin{figure}
   \centering
   \includegraphics[width=0.5\textwidth]{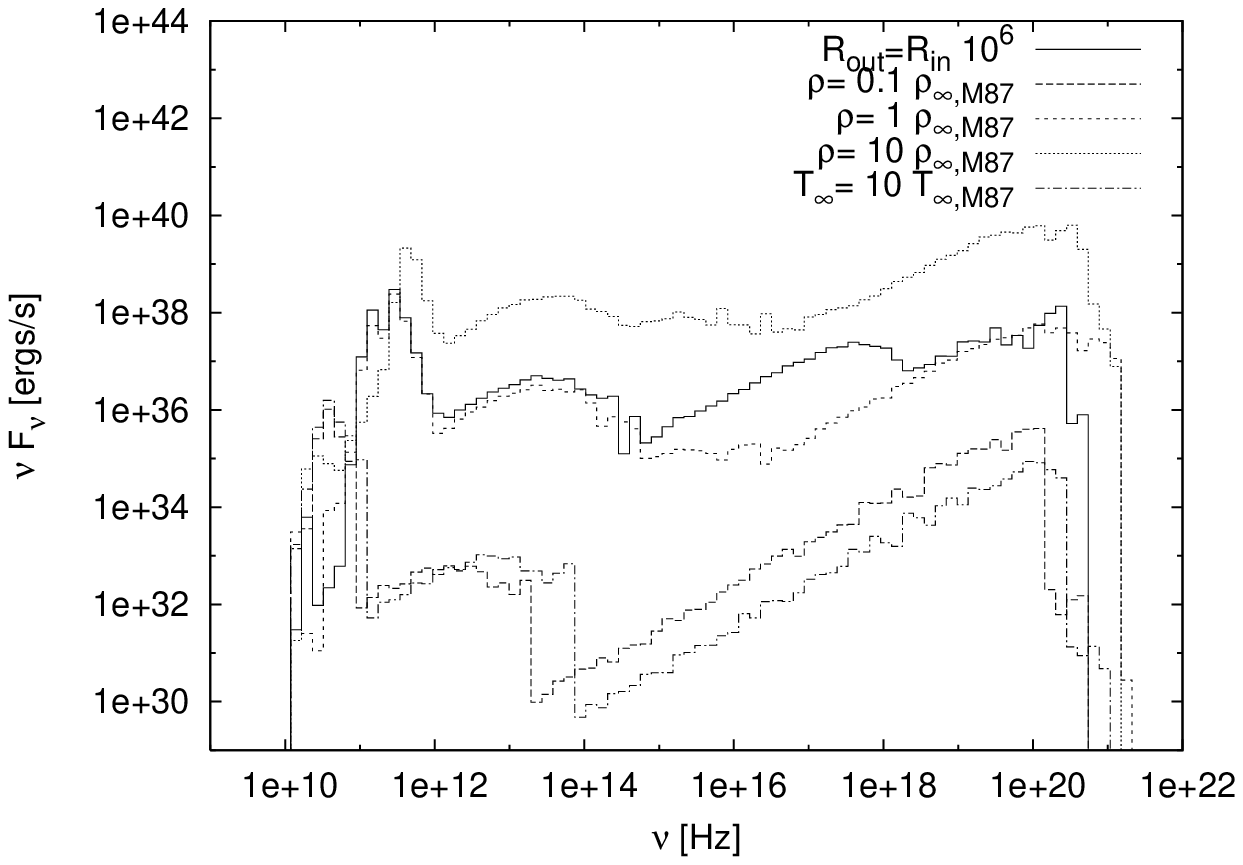}
   \caption{  Spectra for M87 for different boundary conditions. }
   \label{Fig:11}
 \end{figure}

We also checked if external medium conditions have an influence on the 
spectrum formation. The parameters that we change 
are density of the external medium $\rho_{\infty}$ and external medium 
temperature $T_{\infty}$ . By changing these parameters we change 
the structure of the flow and the mass accretion rate. We performed 
calculations for different $\rho_{\infty}$ and fixed other parameter like:
 $T_{\infty}$,$\beta$ and $\delta$. Density profiles for $\rho_{\infty}$= 0.1, 1, 10 $\times 0.36 \cdot 10^{-24} g/cm^3$ (M87 external medium density value),
 are presented in Fig:~\ref{Fig:7}. Electron temperature profiles for these three cases in shown in Fig:~\ref{Fig:8}. The electron temperatures
remain almost the same in this case.   The emitted spectra for these
cases are presented in Fig:~\ref{Fig:11}. X-ray luminosities depends strongly on external medium density. By changing it of order of magnitude,
 we change the mass accretion rate of order of magnitude. As a result the whole spectra (both synchrotron and bremsstrahlung emissivities) 
goes up for higher $\rho_{\infty}$. $log_{10}(L_X)$ for 
$\rho_{\infty}$= 0.1, 1, 10 $\times 0.36 \cdot 10^{-24} g/cm^3$ are respectively 35.02, 37.15, 39.36.
In Fig:~\ref{Fig:9} we also show the effect of waring $T_{\infty}$ (keeping $\rho_{\infty}$ and spherical cloud boundary fixed) 
on density profile. In general density is larger at the
central part of the flow for lower temperatures of the external medium. The electron temperatures convergence to similar values in the center.
As previously emerging radiation spectra are calculated for radius of the scattering cloud equal to Bondi radius. 
The whole radiation spectrum goes down for higher $T_{\infty}$. The effect is stronger for changing $T_{\infty}$, than changing $\rho_{\infty}$. 
We also checked if the adopted outer boundary of the spherical cloud has significant influence on the results. In Fig:~\ref{Fig:11} we
also show spectra of M87 for different boundary radius equal $10^6 \times R_{in}$.  This change affect only the bremsstrahlung part of the spectrum,
which is created by the photons emitted from the outer shells of the spherical cloud. $log_{10}(L_X)$ increase to $log_{10}(L_X=38.24)$, more than one order of magnitude. 
Synchrotron emissivity does not change, because it is created 
by the photons coming very close to the horizon. Small discrepancies are caused by the fact that we divide the flow into N=100 cells in radius,
which gives a certain accuracy of the calculations. The larger boundary radius we take the higher the bremsstrahlung emissivity will be.
\begin{figure}
   \centering
   \includegraphics[angle=0,width=0.5\textwidth]{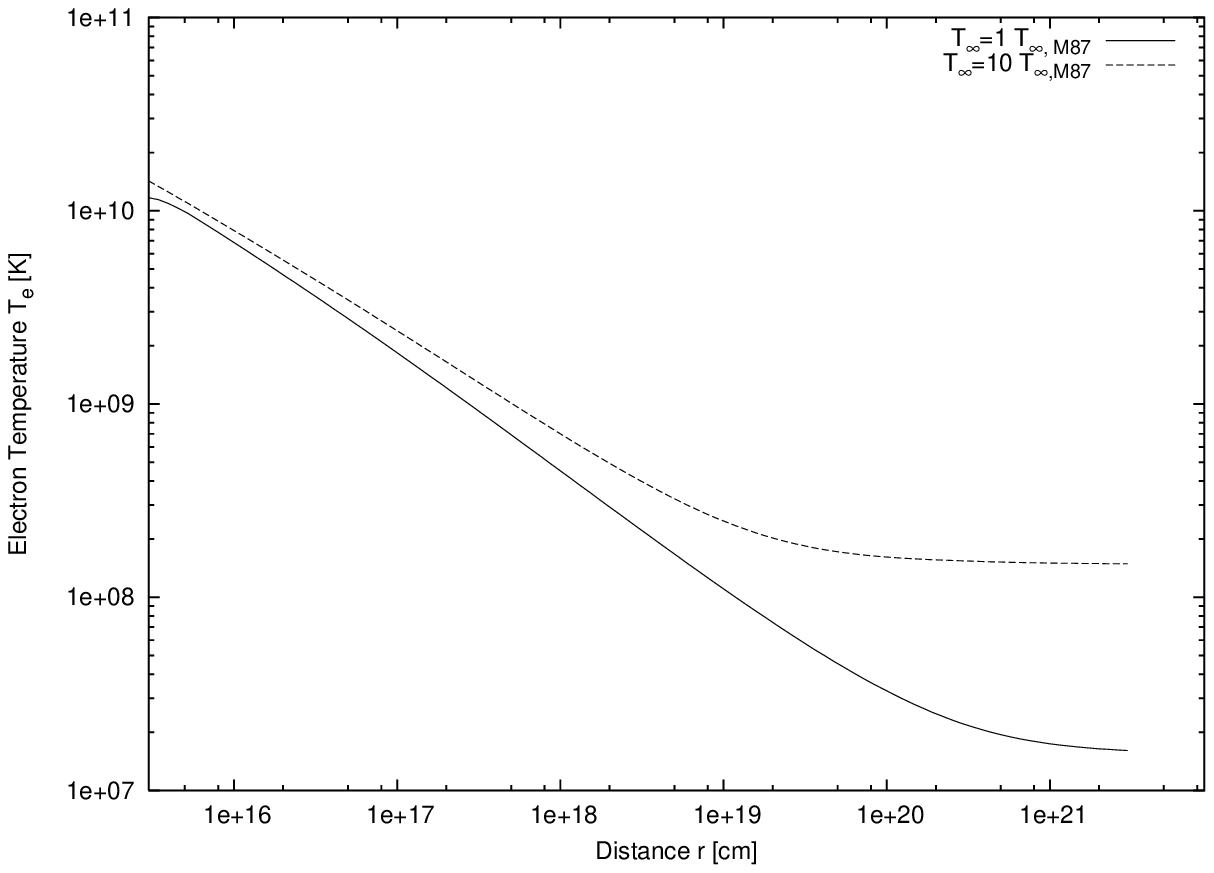}
   \caption{ Electron temperature profiles for different initial $k_B T_{\infty}$. Calculations made for M87.}
   \label{Fig:10}
\end{figure}
\subsection{Emerging spectrum and nuclear luminosities $L_X$ in LLAGN sample}
We calculated spectra for sample of 17 objects collected in P2005 under assumption of the equipartition of the magnetic field
($\beta=0.5$) and assuming basic value $\delta$ = $10^{-3}$ and $\delta=0.5$ for comparison, since the results strongly depends on this parameter. 
The assumed boundary radius of the scattering cloud is assumed to be Bondi radius $R_{Bondi}$, which is proportional to external 
density medium $\rho_{\infty}$ and also depends of $T_{\infty}$. Thus $R_{Bondi}$ is different for each source. 
From obtained spectra we evaluated the expected X-ray luminosities $L_X$ in the 0.3-10 keV band.
The results are presented in Tab.~\ref{tab:1}. In most of the cases modeled X-ray luminosities are much lower
than observed if we assume that $\delta=10^{-3}$. 
In case of NGC221 and NGC5128 the difference is higher than 9 orders of magnitude.
Sources that were under-predicted 4-7 orders of magnitude are: NGC1291, NGC1316, NGC1553, NGC4261, NGC4438, NGC4594, NGC4697 and IC4296.
For NGC4472, M87 and IC1459 the difference between $L_X$ and $L_{X,modeled}$ ranges from 3-4 orders of magnitude.
Sgr A* is under-predicted  about two orders of magnitude.
Good estimation was achieved only for NGC4649 for which observed X-ray luminosities is not larger than 
order of magnitude from the modeled one. 
Three limitation values in case of NGC821, NGC1399, NGC4636 are not exceeded.

\begin{table*}

\centering
\begin{tabular}{c c c c c c c c c c}      
\hline 
Source &$M_{BH}$            &$kT_{\infty}$&$\rho_{\infty}$        & $\dot{m}^x$ & observed        & cal. $L_{X,nucl}$**& $\eta$***      &cal. $L_{X,nucl}$** &$\eta$***\\
       & $(10^8 M_{\odot})$ &(keV)        & $(10^-{24} gcm^{-3})$ &             &  $L_{X,nucl}$ * &  $\delta=0.001$    & $\delta=0.001$  & $\delta=0.5$      & $\delta=0.5$\\

\hline                    
     NGC221(M32) & 0.025 & 0.37 & 0.13 &  $3.1 \cdot 10^{-7}$ &  36.44      &  27 &  $9.8 \cdot 10^{-9}$ &33.89& $2.4 \cdot 10^{-3}$\\
     NGC821      & 0.37  & 0.46 & 0.01 &  $2.5 \cdot 10^{-7}$ & $\le 38.66$ &  28.81 & $2.9 \cdot 10^{-9}$ & 33.41 & $3.1 \cdot 10^{-4}$\\
     NGC1291     & 1.1   & 0.34 & 0.56 &  $6.7 \cdot 10^{-5}$ &39.60        &  34.26 & $5.56\cdot 10^{-7}$ & 38.78 & $1.9 \cdot 10^{-3}$\\
     NGC1316     & 3.9   & 0.62 & 0.44 &  $7.6 \cdot 10^{-5}$ &38.87        &  34.76 &   $1.38\cdot 10^{-7}$ & 38.3 & $1.6 \cdot 10^{-4}$\\
     NGC1399     & 12.0  & 0.8	& 0.47 &  $1.6 \cdot 10^{-4}$ & $\le39.14$  &  35.94 &   $ 1.9 \cdot 10^{-7}$ & 40.33 &  $1.9 \cdot 10^{-3}$\\
     NGC1553     & 1.6   & 0.51 & 0.6  &  $5.4 \cdot 10^{-6}$ &40.01        &  34.15  & $1.47 \cdot 10^{-7}$  &  37.99 & $0.015$\\
     NGC4261     & 5.4   & 0.6  & 0.17 &  $4.2 \cdot 10^{-5}$ &41.15  &  35.08 & $1.41 \cdot 10^{-7} $ & 39.11 &  $3.9 \cdot 10^{-3}$\\
     NGC4438     & 0.5   & 0.58 & 0.99 &  $2.4 \cdot 10^{-6}$ &39.65  &  32.93 & $1.01\cdot 10^{-7}$& 38.39 &  $0.02$\\
     NGC4472     & 7.9   & 0.8  & 0.32 &  $5.6 \cdot 10^{-5}$ &38.69  &  35.10 &$ 1.74 \cdot 10^{-7}$ & 39.31&  $2.7 \cdot 10^{-3}$\\
     NGC4468(M87)& 34.0  & 0.8  & 0.36 &  $3.3 \cdot 10^{-4}$ &40.88  &  37.13 & $2.92 \cdot 10^{-7}$ & 41.15 &  $1.6 \cdot 10^{-3}$\\
     NGC4594     & 10.0  & 0.65 & 0.29 &  $8.7 \cdot 10^{-5}$ &40.34  &  35.6 &$ 1.5 \cdot 10^{-7}$& 40.28& $2.9 \cdot 10^{-3}$\\
     NGC4636     & 3.0   & 0.6  & 0.11 &  $1.3 \cdot 10^{-5}$ & $\le38.41$ &  33.2 & $ 4.6 \cdot 10^{-8}$& 38.44& $3.0 \cdot 10^{-3}$ \\ 
     NGC4649     & 20.0  & 0.86 & 1.05 &  $8.1 \cdot 10^{-4}$ &38.11  &  37.29 &$ 6.16 \cdot 10^{-7}$ &41.37 & $4.9 \cdot 10^{-4}$\\
     NGC4697     & 1.7   & 0.33 & 0.05 &  $9.7 \cdot 10^{-6}$ &38.64  &  32.65 &$2.79 \cdot 10^{-8} $ & 37.62& $2.1 \cdot 10^{-3}$\\
     NGC5128     & 2.4   & 0.5  & 0.08 &  $1.0 \cdot 10^{-5}$ &42.11  &  32.9 &$3.44 \cdot 10^{-8}$ & 38.21 & $3.0 \cdot 10^{-3}$\\
     IC1459      & 25.0  & 0.5  & 0.54 &  $7.0 \cdot 10^{-4}$ &41.18  &  37.7 & $7.95 \cdot 10^{-7}$& 42.25 & $4.9 \cdot 10^{-5}$\\
     IC4296      & 11.0  & 0.56 & 1.0  &  $6.7 \cdot 10^{-4}$ &41.38  &  36.99 & $6.1 \cdot 10^{-7}$ & 41.16&$7.0 \cdot 10^{-4}$\\
     GC          & 0.037 & 1.3  & 52.0 &  $2.3 \cdot 10^{-5}$ &33.38  &  31.21 &  $1.1 \cdot 10^{-7}$  & 37.57 & $4.7 \cdot 10^{-4}$\\ 
\hline                  
\end{tabular}
 \caption{Results of the simulation assuming the equipartition of the magnetic field ($\beta=0.5$), $L_X$ refer to a 0.3-10 keV band.
Most of the sources are much more brighter in X-rays than predicted by our model. Very good estimation of $L_X$ was obtained for
 NGC4649. This galaxy is elliptical. Rest of the predicted X-ray luminosities are 3-9 orders of magnitude larger than observed.
For parameter $\delta=0.5$ the results are different. Most of the sources is accomodated to the observations. 
($*$) for references see P2005,($**$) our estimation of $L_X$ emerging from Bondi accretion flow. $\eta^{***}$ calculated efficiency of 
spherical accretion defined as $\eta=L_{bol}/ \dot{M} c^2$.(x) $\dot{m}=\dot{M}/\dot{M}_{Edd}$,$\dot{M}_{Edd}=L_{Edd}/0.1c^2$}          
\label{tab:1}
\end{table*}

\begin{table*}
\centering  
\begin{tabular}{c c c}      
\hline 
Source  & features & ref\\
\hline                    
     NGC221(M32) & elliptical dwarf, radio emission & (1)  \\
     NGC821      & elliptical  & (2) \\
     NGC1291     & early type spiral & (3)\\
     NGC1316     & disturbed early type elliptical, giant radio lobes, S-shaped nuclear radio-jets & (4)\\
     NGC1399     & elliptical & (5)\\
     NGC1553     & spiral feature passing threw the center, weak radio source & (6)\\ 
     NGC4261     & elliptical radio galaxy, kiloparces scale jets, bright nuclear optical source surrounded by a disk of gas and dust & (7)  \\
     NGC4438     & spherical bulge, outflow bubbles in Northwest-south east direction from the nucleus & (8) \\
     NGC4472     & giant elliptical galaxy,  cavities (inner 2') corresponding to the position of radio lobes &\\
                 & two small extended sources within 10'' of the nucleus of the galaxy, both the same luminosity & (9)\\                                               
     NGC4468(M87)& elliptical radio galaxy, one-sided jet, large radio structure & (10)\\
     NGC4594     & spiral SAa type & (11)\\
     NGC4636     & elliptical, 'spiral arm in the core, cavity presence, hot gas violently disturbed & (12) \\ 
     NGC4649     & elliptical & (13)\\
     NGC4697     & early type elliptical & (14)\\
     NGC5128(Cen A)     & AGN,giant elliptical radio galaxy, inner jet, radio-lobe & (15)\\
     IC1459      & extremely radio-laud compared to normal radio-laud Quasars, &\\
                 & giant elliptical galaxy, counter rotating core (merger possible) & (16)\\
     IC4296      & giant elliptical radio galaxy, large-scale jets & (17)\\
     GC          & shows flaring & (18)\\ 
\hline                  
\end{tabular}
\caption{ Features of the galaxies in a calculated sample. (1) Ho et al. 2003a, (2) Fabbiano et al. 2004, (3) Irwin et al. 2002, (4) Dong-Woo Kim et al. 2002, 
(5) Boute et al. 2002, (6) Blauton et al. 2001, (7) Zezas et al. 2005, (8) Machacek et al. 2004, (9) Biller et al. 2004, (10) Di Matteo et al. 2003, 
(11) Pellegrini et al. 2003, (12) O'Sullivan et al. 2005, (13) Randall et al. 2004, (14) Sarazin et al. 2001, (15) Evans et al. 2004, (16) Fabbiano et al. 2003,
(17) Pellegrini et al.(2003a)  (18) Yuan et al. 2002, Senda et al.} 
 \label{tab:2} \end{table*}
In  Tab.~\ref{tab:2} we collect the characteristic properties of the studied sample of galaxies, to see whether 
the estimated values can be modified including some additional effects. 
In an extreme case of NGC5128(Cen A) where emitted X-ray luminosity differ of more than 9 orders of magnitude, 
our model is not appropriate. In this source calculated efficiency of accretion is $\eta=L_{X,observed}/ \dot{M}_{Bondi} c^2=0.32$. 
Because NGC5128(Cen A) is an active galactic nucleus, given by P2005 for comparison, the result is reliable.
The other source with difference of modeled $L_X$ and observed one larger that 9 orders of magnitude is an elliptical dwarf galaxy with
observed radio emission.
In case of sources for which $L_X/L_{X,modeled}$ is between 4 and 7 orders of magnitude, observations may indicate 
for existence of angular momentum 
inside the core. NGC1291 is an early type spiral, NGC1316 is disturbed early type elliptical,
 NGC 1553 has a spiral feature passing threw a center, NGC 4261 core is surrounded by a disk
 of gas and dust, NGC4438 has an outflow bubbles which also may indicate angular momentum
to dominate the accretion flow, NGC4594 is a spiral galaxy, the last two sourced NGC 4697
 and IC4296 are elliptical radio galaxies.
Sources where $L_X/L_{X,modeled}$ is between 3 and 4 orders of magnitude namely NGC4472 and
 M87 are ellipticals radio galaxies with 
jets or radio lobes. NGC4649 which is modeled very well by our calculations is an elliptical
 galaxy. The sources for which we have only limitation
values of X-ray emission are ellipticals. In NGC4636 we see 'spiral arm in the core'. \\
For $\delta=0.5$ the results change dramatically. The efficiency of accretion grows 3, even 5 orders.
One source: Galactic Center differ more than 4 orders of magnitude from its observational values. 
For Galactic Center the luminosity is over-predicted. The differences between modeled and measure $L_X$ lay in range 2-4 orders of magnitude
for NGC 1553, NGC4261, NGC4649 and NGC 5128 (AGN).
NGC4649 is overpredicted more than 3 orders of magnitude. Modeled luminosity of NGC5128 is too low in comparision with observations.
Two sources: NGC1553 and NGC 4261 are still under-predicted of about 2 orders of magnitude.
Other sources (NGC1291, NGC1316, NGC4438, NGC4472, M87, NGC4594, NGC4697, IC1459, IC4296) 
differ from their observational counterparts of about 1 order of magnitude, lots of them even less than order of magnitude (see Table 1.)
Limitation value for NGC821 is not exceeded, and it is still possibly 5 orders of magnitude under-predicted. In case of NGC1399 
modeled luminosity is order of magnitude too high than limitation value. In case of NGC 4636 is also exceeded, but it is very close to 
its limit. 
The source that fit well to observations for $\delta$=0.001, NGC4649, is now more than 3 orders of magnitude over-predicted.\\

\section{Discussion and conclusions}
In this paper we reconsider the sample of 17 LLAGNs (with measured 
temperature and density of surrounding medium) and GC collected
in paper of P2005. 
Usually LLAGN are discussed in context of ADAF model (Quataert 1999, 
Di Matteo et al. 1999, Di Matteo et al. 2003, Loewenstein et al. 2001). In general ADAF 
over-predicts measured luminosities, if one assume that the mass accretion
rate of the flow is equal to Bondi mass accretion rate. As it is pointed
out by Narayan (2002) (also Quataert 2003), one should include into calculation 
mass accretion rate reduced by the $\alpha$ factor (dimensionless viscosity parameter)
 so that $\dot{M}_{ADAF}=\alpha \dot{M}_{Bondi}$, to make the model self-consistent.
Although ADAF model, with proper assumption of $\dot{M}$, predicts well X-ray luminosities
in some cases, it can over-predicts emission in radio band (Loewenstein 2001). Also 
in case of the sources with jet structures, it is hard to include additional emission
in radio band (Narayan 2002). The second point concerns estimation of the Bondi mass 
accretion rate $\dot{M}_{Bondi}$. $\dot{M}_{Bondi}$ may be an order of magnitude smaller depending
on polytropic index $\gamma$ and mean particles mass $\mu$. E.g. for NGC1399 the mass accretion rate 
assuming $\gamma=5/3$ and $\mu=0.5$ is $\dot{M}_{Bondi}=4.6 \cdot 10^{-3} M_{\odot}/yr$, but if we assume different
$\gamma=1.4$ (which is typical for the partially ionized ISM and usually assumed in ADAF models) and different $\mu=1.0$
we obtain an order of magnitude higher mass accretion rate $\dot{M}_{Bondi}=4.2 \cdot 10^{-2} M_{\odot}/yr$.
One should pay a particular attention in choosing constants determining accretion rate, because emerging spectrum
depends strongly on this quantity. In our calculations we take $\gamma \approx 5/3$ and $\mu=0.5$. 

Narayan (2002) also points out that LLAGN weak luminosities can be be fitted to the data assuming two-temperature Bondi model.
In this paper we apply a spherically symmetrical Bondi model of accretion to the sample of sources from P2005, and we
calculate the radiation spectra emerging from accreting plasma, including a full treatment of radiation transfer through the gas. 
Historically, the problem of spherical accretion onto a compact
objects was present in many papers (for review see Nobili et al. 1991). Although model of steady spherical flow
is only a mathematical model (not realized in nature because there is always some angular momentum present), it can
be used as a rough approximation of an accretion with very small angular momentum (as we would expect to be present in elliptical galaxies). 
The similar model of spherical accretion was tested 
by Yim and Park (1995) in case of our Galactic Center. 
(Detailed comparison to Yim and Park(1995) is not possible because the paper is unavailable).
Melia (1994) considered semi-spherical flow (with very low angular momentum) but with the disc inside.
 Authors in both papers conclude that spectrum can be reasonably explained
 with such assumptions. 

Our calculations also show that spherical accretion can reconstruct observed X-ray luminosities 
in most cases of the LLAGN sources presented in P2005, within an order of magnitude error (10 cases in 17 galaxies in a sample), for parameter $\delta$=0.5. 
The nuclear luminosities $L_X$ are very sensitive on the changes of the 
heating parameter $\delta$.
This parameter indicate how much accretion energy will be transfered to the electrons, and  determine the electron temperature profile.
($\beta$ parameter does not influence the electron temperature significantly) 
For most of the sources X-ray luminosities are strongly 
under-predicted when $\delta=0.001$. The obtained efficiency $\eta$ of spherical accretion in all cases for this value of $\delta$ 
is very small, of the order of $10^{-7}-10^{-9}$. If we assume $\delta$=0.5, 
the electrons are heated in the same degree as ions. Half of the accretion energy is transfered to ions, the other half to the electrons.
But the increase in X-ray luminosity is not caused by the growth of bremsstrahlung emissivity. 
When the temperature grow, the bremsstrahlung
emission goes up like $T^{0.5}$ and it is more sensitive to the changes of density (like $\rho^2$), than to the electrons temperature. 
So the effect if not very strong for the emissivity of bremsstrahlung radiation.
The reason of increase of $L_X$ is that the synchrotron emissivities grows, and X-ray part of the spectrum 
is dominated by the Compton scattered synchrotron photons.
The efficiency grows even 5 orders of magnitude for $\delta$=0.5.
Our calculations show that the assumption, that electrons are heated directly by accretion in the same degree as ions, is
more proper that assuming $\delta$ to be 0.001 because of mass ratio ($m_e/m_p$) (e.g. Esin et al. 1997, Manmoto 2000)  
This is in agreement with the results of Bisnovatyi-Kogan $\&$ Lovelace (1997). The authors argue that in a presence of magnetic field and 
plasma instabilities, gravitational energy is transfered predominantly to electrons. Also in work of Yuan (et al.2003) the assumption of $\delta=0.55$
allows to accommodate the spectrum of Sgr A* with ADAF model. 

Bremsstrahlung emissivity
depends strongly on external medium conditions ($T_{\infty}$ and $\rho_{\infty}$) which control the mass accretion rate. 
Values of these quantities are estimated by observations, so they rather cannot be changed significantly in this approach.
  $L_X$ in case of over-predicted sources could be reduced and accommodated to observation with assumption of 
an accretion with outflows. It was considered by e.g. Quataert $\&$ Narayan (1999). For one of the sources 
(Galactic Center), nuclear luminosity $L_X$ was strongly over-predicted by our model for $\delta=0.5$.  
External medium measurements indicate (assuming Bondi flow) mass accretion rate for this source to be $\sim 10^{-6} \rm{M_{\odot}/yr}$. 
On the other hand, we have also mass accretion rate limitations estimated by measurements of polarization of
radiation near the black hole in Sgr A*. From Faraday rotation we obtain limit for 
$\rm{\dot{M}} \sim 10^{-7} \rm{M_{\odot}/yr}$ (Atiken et al. 2000).
This fact also additionally eliminate simple ADAF model for this source (Bower et al. 2005).
Differences between $\rm{\dot{M}}$ near the capture radius and $\rm{\dot{M}}$ near horizon
indicate that most of the accreting material do not reach a black hole. For 
Sgr A* the possible explanation could be the accretion flow disturbed by the outflow. 
 For Sgr A*
the radio part of a spectrum is too high in our model for $\delta=0.5$,
 also it is higher than NIR limitations for mass
 accretion rate $\dot{M} \sim 10^{-6} \rm{M_{\odot}/yr}$. The outflow could 
be a possible explanation of spectrum and also reduce mass accretion rate.
For $\delta=0.001$ the emission was very low, the peak of the synchrotron
 radiation attain in $\nu F_{\nu}=10^{34}$ [ergs/s]. ADAF model in paper 
of  Narayan (et al. 1998) gives a more denser accretion flow than Bondi flow,
 since the radial velocity is smaller. Thus our results  for $\delta$=0.001 are much below the observational points.
In Yuan (et al. 2003) the emission is accommodated 
to X-ray emission also in assumption of higher $\delta=0.55$ value  but for  $\dot{M}=10^{-8} M_{\odot}/yr$.
 The reason why our model for Sgr A* gives so poor constrains for $\delta=0.5$ 
in comparison to Yuan (et al. 2003)
is that we assume Bondi accretion rate which is two orders of magnitude
larger than  $\dot{M}=10^{-8} M_{\odot}/yr$.
Our model can reproduce better the Sgr A* spectrum if we allow for the
specific adjustment of the parameters like $\dot M$ and $\delta$ since the
synchrotron emissivity is very sensitive to the temperature changes
(controlled mainly by these two parameters). Additionally, our model can also
explain the soft spectral slope seen in the data (Baganoff et al. 2003) since
the comptonized synchrotron component can extend to the X-ray band. However,
the model never fits the radio frequencies below $\sim 10^{11}$ Hz. This part
of the spectrum can be only explained by models which include non-thermal
populations of electrons, like e.g. ADAF-jet model of Yuan et al. (2002). The
presence of a jet-like outflow also in other sources is supported by observations (see Tab.~2).
Also the negligence of the spherically symmetric outflow is a serious weakness
of our model since there are now observational evidences for Sgr A* that accretion rate
is not constant with radius (e.g.Bower et al. 2003).

The effect of the angular momentum of the accretion flow also cannot be neglected
(the low angular momentum was considered by Proga et al. 2003, but without any estimations of emerging spectrum).
Bondi spherical accretion model is a mathematical model. In reality there is always angular momentum.
We also know that standing shocks are the part of a low angular momentum flows (Das 2002, Das, Pendharkar $\&$ Mitra 2003).
Also tubulent flow may create shocks.
  In our model we assume the thermal distribution of electrons. This is a weakness of the model, since the shock 
produce some fraction of non-thermal electrons.
Thanks to studies of Sgr A* (Mahadevan 1999, Ozel et al. 2000, Yuan et al. 2003) with hybrid (thermal + non-thermal power-law tail) population of electrons we know that
observed low-frequency radio spectrum of Sgr A* can be explained if small fraction of electrons is non-thermal. 
Close to a black hole turbulence and magnetic reconnection can accelerate electrons. Synchrotron emission from this electrons and 
Compton scattering on them can be a reason of flaring in broadband spectrum.
In all calculations of spectra for M87 the low-frequency radio emission wasn't reconstructed, this may be a 
reason why non-thermal electrons should be included into calculations, moreover scatterred nonthermal photons can affect the X-ray band of the spectrum.

Our model is also not consistent with recent results of modeling spherical accretion with magnetic fields.
Because we include synchrotron emission in calculating the radiation spectrum we assume that there is 
some random magnetic field in plasma surrounding a black hole. Spherical flows including 
magnetic field are unstable (e.g. Igumenshchev et al. 2002), thus we are able to obtain only some mean luminosity
of spherical accretion. Our model is not able to reconstruct any detailed features of time-dependent spectrum 
like e.g. flares. Another disadvantage of our model is that the equations of spherically symmetrical 
flow are calculated in Newtonian regime. Including relativistic effects may be 
important very close to a horizon of a black hole, 
where in our calculations most of the synchrotron radiation comes from.

\begin{acknowledgements}
We would like to thank Bozena Czerny, Aneta Siemiginowska, 
Piotr Zycki and Agata Rozanska for helpful comments and
useful discussions.
Part of this work was supported by grant
PBZ-KBN-054/P03/2001 and 1P03D00829 of the Polish State Committee for Scientific
Research.

\end{acknowledgements}

\end{document}